\documentclass[aps,prl,reprint,superscriptaddress,longbibliography]{revtex4-2}

\usepackage[dvips]{graphicx}
\usepackage{hyperref}
\usepackage{xcolor,soul}
\usepackage{amsmath}

\begin{document}

\title{A self-calibrating superconducting pair-breaking detector}

\author{E. T. Mannila}
\email{elsa.mannila@aalto.fi}
\affiliation{QTF Centre of Excellence, Department of Applied Physics, Aalto University, FI-00076 Aalto, Finland}

\author{V. F. Maisi}
\affiliation{Physics Department and NanoLund, Lund University, Box 118, 22100 Lund, Sweden}

\author{J. P. Pekola}
\affiliation{QTF Centre of Excellence, Department of Applied Physics, Aalto University, FI-00076 Aalto, Finland}

\date{\today}

\begin{abstract}
We propose and experimentally demonstrate a self-calibrating detector of Cooper pair depairing in a superconductor based on a mesoscopic superconducting island coupled to normal metal leads. On average, exactly one electron passes through the device per broken Cooper pair, independent of the absorber volume, device or material parameters. The device operation is explained by a simple analytical model and verified with numerical simulations in quantitative agreement with experiment. In a proof-of-concept experiment, we use such a detector to measure the high-frequency phonons generated by another, electrically decoupled superconducting island, with a measurable signal resulting from less than 10 fW of dissipated power.
\end{abstract}

\maketitle

\paragraph{Introduction.}
A key prediction of the Bardeen-Cooper-Schrieffer theory of superconductivity is the existence of an energy gap $\Delta$ for single-particle excitations. 
The resulting exponentially suppressed density of thermal excitations makes superconductors very sensitive to radiation at frequencies higher than $2\Delta/h$ with $h$ the Planck constant, which, although detrimental for superconducting circuits used in quantum computing \cite{martinis2009energy, catelani2011quasiparticle, corcoles2011protecting}, enables applications as detectors.
Pair-breaking superconducting detectors, such as superconducting tunnel junction \cite{lerch2005quantum}, kinetic inductance \cite{day2003broadband,zmuidzinas2012superconducting} and quantum capacitance \cite{echternach2018single} detectors have found use in physics and astronomy, enabling single-photon detection at optical \cite{peacock1996single} and terahertz \cite{echternach2018single} frequencies. They can also be used as phonon-mediated detectors \cite{swenson2010highspeed, moore2012position}, and tunnel junction detectors have been used for phonon spectroscopy \cite{bron1980spectroscopy, wybourne1988phonon, otelaja2013design, hertzberg2014direct}.

In all of these devices, inferring the number of broken Cooper pairs from the measured response requires calibration or modeling. 
In contrast, in this Letter we present a mesoscopic pair-breaking detector whose response is given simply by a current 
\begin{equation}\label{eq:i}
I = e\Gamma_{\text{pb}},
\end{equation}
where $e$ is the elementary charge and $\Gamma_{\text{pb}}$ is the rate at which Cooper pairs are broken.
We describe the device operation with a simple analytical model, which agrees with the predictions of a full numerical model in quantitative agreement with experiment. 
In our proof-of-concept experiment, we measure the response of the detector to pair-breaking phonons emitted by another superconducting island, while ruling out that the response could be due to non-pair-breaking mechanisms by comparison with a normal-metallic reference source. 
We extract the fraction of phonons transmitted from emitter to detector 
and find that dissipated power as low as 10 fW in the emitter is enough to create a measurable signal in the detector. 
Although the number of Cooper pairs generated per absorbed phonon or photon depends on the frequency \cite{guruswamy2014quasiparticle}, due to its well-defined absorption volume and self-calibrating operation we foresee our device as particularly useful for studying propagation of athermal phonons. This is important in detectors \cite{henriques2019phonon, karatsu2019mitigation, martinez2019measurements} as well as applications in quantum information based on superconducting circuits, where phonons may cause quasiparticle poisoning over large distances \cite{patel2017phononmediated, wilen2020correlated}.

\paragraph{Operating principle.}

\begin{figure}
\includegraphics{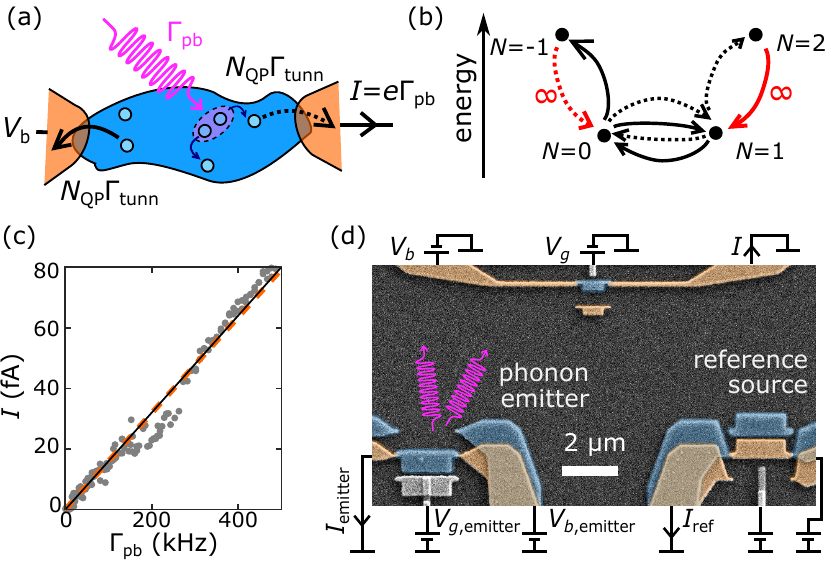}
\caption{\label{fig:sketch}
(a) Sketch of device. Cooper pairs are broken when radiation is absorbed in a superconducting island, and the resulting quasiparticles decay by tunneling to the normal metal leads (black arrows) with a rate $N_{\text{QP}} \Gamma_{\text{tunn}}$ for $N_{\text{QP}}$ excitations on the island. 
(b) Energies of states with $N=-1,0,1,2$ excess electrons compared to charge neutrality on the island, when the gate offset is close to $n_g = 0.5$. Black arrows indicate transitions which remove a quasiparticle through the left (solid lines) or right (dashed lines) tunnel junction, while red arrows indicate transitions which add a quasiparticle into the island. The quasiparticle population is set by transitions between charge states $N=0,1$, while net current flows via cycles involving the excited charge states $N=-1, +2$. 
(c) Numerical verification of operation. The ideal response $I=e\Gamma_{\text{pb}}$ (solid line) is reproduced by our simulations (dashed line) within 1\% up to $\Gamma_{\text{pb}}$ = 500 kHz. Symbols show the experimental response, where $\Gamma_\text{pb}$ is linearly proportional to the measured $I_\text{emitter}$.
(d) Proof-of-concept experiment and sketch of measurement setup, with pair-breaking detector (top), superconducting phonon emitter (bottom left) and normal-metallic reference source (bottom right). 
}
\end{figure}

Our device, a superconducting island with charging energy $E_C$ smaller than the superconducting gap $\Delta$, is sketched in Fig. \ref{fig:sketch}(a). We operate at low temperatures $k_B T \ll E_C, \Delta$ such that the probability of thermally excited quasiparticles is negligible. Incident radiation (pink) breaks Cooper pairs in a mesoscopic superconducting aluminum island (blue) at a rate $\Gamma_{\text{pb}}$. The resulting quasiparticle excitations (light blue circles) relax by tunneling to the normal metal leads through the left or right tunnel junctions with a rate $N_{\text{QP}}\Gamma_{\text{tunn}}$ for $N_{\text{QP}}$ excitations on the island. Since relaxation can happen equally likely through either junction, this process carries no current, but sets the time-averaged quasiparticle population to 
\begin{equation}\label{eq:qp-population}
\langle N_{\text{QP}} \rangle = \Gamma_{\text{pb}} / \Gamma_{\text{tunn}}.
\end{equation} 
As quasiparticles can tunnel out as both electrons and holes, the tunneling events occur between the lowest-energy charge states with $N= 0,1$ excess electrons on the island, when the gate offset is close to charge degeneracy, $n_g \approx 0.5$. 
In our model, the energy cost $\Delta$ of creating quasiparticles is accounted for by explicitly tracking the number of quasiparticles $N_{\text{QP}}$ on the island \cite{maisi2013excitation}.

In the presence of a small applied bias voltage $V_b < \Delta/e$, quasiparticles can also tunnel out to the higher-energy charge states $N=-1$ ($2$) through the right (left) tunnel junction with the same rate $N_{\text{QP}} \Gamma_{\text{tunn}}$, such that charge is transported in the direction of the bias in both cases.
Due to the high energy cost of occupying these states, the island will then return to its previous charge state near-instantaneously, when a new quasiparticle tunnels in through the left (right) tunnel junction. This occurs again in the direction of the bias, as indicated by the red arrows in Fig. \ref{fig:sketch}(b), leading to one net electron transported through the device. 
The cycles $N=0 \rightarrow -1 \rightarrow 0$ and $N = 1 \rightarrow 2 \rightarrow 1$ do not change the quasiparticle population on the island, yet they determine the current through the device.
The total current is set by the rate of the first step of the cycles $N_{\text{QP}} \Gamma_{\text{tunn}}$. Combined with Eq. \eqref{eq:qp-population}, we obtain the result of Eq. \eqref{eq:i}: $I = e \langle N_{\text{QP}} \rangle \Gamma_{\text{tunn}} = e \Gamma_{\text{pb}}$. 
Here, we have assumed equal tunneling rates in both junctions for simplicity, but the result holds for unequal junctions as well \footnote{See Supplemental Material, which includes Refs. \cite{pekola2010environment, taupin2016inas, higginbotham2015parity, karimi2020optimized, zorin1995thermocoax, giazotto2006opportunities, kautz1993selfheating, kauppinen1996electronphonon, meschke2004electron, wang2019crossover, maisi2011realtime, pekola1994thermometry, savin2006thermal, feshchenko2017thermal, chang1977kineticequation, rothwarf1967measurement, averin1990virtual, averin1992singleelectron, knowles2012probing, marinsuarez2020active, kaplan1976quasiparticle, kaplan1979acoustic, oneil2012measurement, mrzyglod1994mean, bobetic1964evaluation}, for additional derivations, details on the numerical simulations, fabrication and measurements, normal-state characterization of the device, and modeling of phonon emission}. \nocite{pekola2010environment, taupin2016inas, higginbotham2015parity, karimi2020optimized, zorin1995thermocoax, giazotto2006opportunities, kautz1993selfheating, kauppinen1996electronphonon, meschke2004electron, wang2019crossover, maisi2011realtime, pekola1994thermometry, savin2006thermal, feshchenko2017thermal, chang1977kineticequation, rothwarf1967measurement, averin1990virtual, averin1992singleelectron, knowles2012probing, marinsuarez2020active, kaplan1976quasiparticle, kaplan1979acoustic, oneil2012measurement, mrzyglod1994mean, bobetic1964evaluation}
Because the quasiparticle tunneling rates are constant over a range of energy \cite{saira2012vanishing}, the current due to pair-breaking forms diamond-shaped plateaus as a function of $V_b$ and $n_g$. Although the size and location of these plateaus depends on $E_C$, they exist for all $E_C < \Delta$ \cite{Note1} and neither $E_C$, $V_b$ nor $n_g$ need to be tuned precisely for the self-calibrating operation.

We validate this simple picture by performing numerical simulations based on a rate equation tracking the occupation probabilities of states with $N$ excess electrons and $N_{\text{QP}}$ quasiparticle excitations on the island \cite{maisi2013excitation}. The simulations incorporate single-electron and Andreev tunneling at finite temperature of the normal metal leads, as well as quasiparticle recombination through the electron-phonon coupling \cite{maisi2013excitation,Note1}. 
We find that with our device parameters, 
the effect of finite temperature and Andreev tunneling is negligible around $n_g = 0.5$ and $|V_b|<150~\mu$V. The non-zero electron-phonon recombination rate, scaling as $\Gamma_{\text{R}} N_{\text{QP}}^2$ with the prefactor $\Gamma_{\text{R}}$ depending on the device parameters, reduces the quasiparticle population from the value of Eq. \eqref{eq:qp-population} and the current response. The condition that recombination be negligible compared to relaxation by tunneling, $\Gamma_{\text{R}} \langle N_{\text{QP}} \rangle^2 \ll \Gamma_{\text{tunn}} \langle N_{\text{QP}} \rangle$, can be expressed as 
\begin{equation}\label{eq:recombinationlimit}
\Gamma_{\text{pb}} \ll  \frac{12 \zeta (5) k_B^5}{\Sigma  \mathcal{V} \Delta^2 e^4 R_T^2},
\end{equation}
where $\Sigma$ is the electron-phonon coupling constant, $\mathcal{V}$ is the absorber volume, $\zeta$ is the Riemann zeta function, and $k_B$ is the Boltzmann constant \cite{Note1}. 
For our device parameters, the right-hand side evaluates to 70 MHz, corresponding to femtowatts of absorbed power. The simulated current is within 1\% of the ideal value up to 500 kHz, the value of $\Gamma_{\text{pb}}$ reached in the experiment, as shown in Fig. \ref{fig:sketch}(c). 

\paragraph{Proof-of-concept experiment.}

\begin{figure*}
\includegraphics{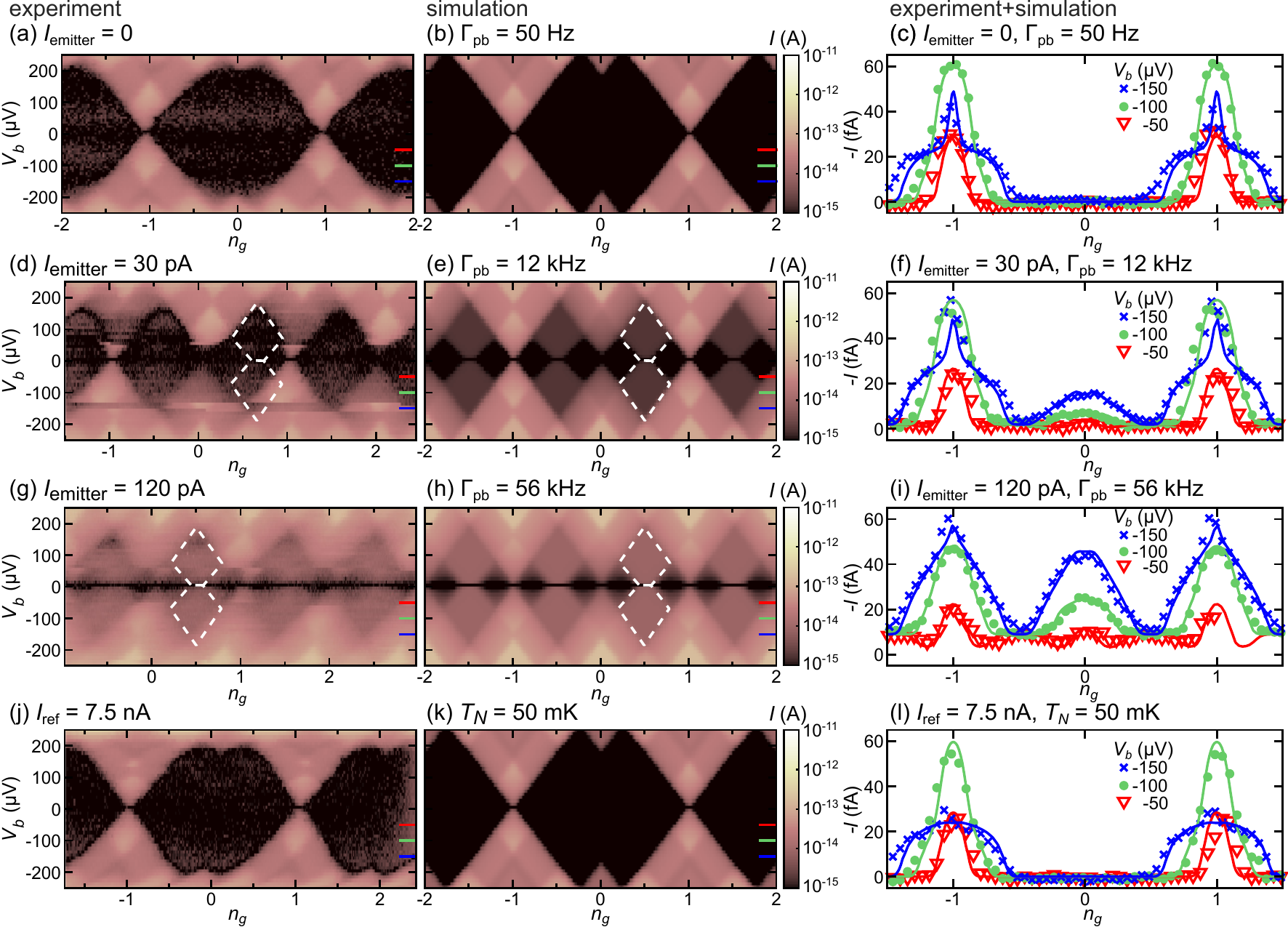}
\caption{\label{fig:pinkabsorber}
Measured (left column and symbols in right column) and simulated (middle column and solid lines in right column) subgap current through our pair-breaking detector. On the first row, no current is applied to the emitter or reference source, while the current increases substantially when the current through the phonon emitter is increased to 30 pA (second row) or 120 pA (third row). This is reproduced quantitatively in the simulations by changing only the rate of pair-breaking radiation $\Gamma_{\text{pb}}$. In contrast, increasing the current through the reference emitter up to 7.5 nA (bottom row) creates nearly no change in the detector response, proving that the response is indeed due to pair-breaking phonons. White dashed lines in panels (d,e,g,h) indicate the regions where the self-calibrating response is expected. The right-most column shows cuts in the data at $V_b=-50~\mu$V, $-100~\mu$V and $-150~\mu$V, also indicated by colored lines in the left and middle columns.  %In panels (d) and (g), there are several jumps of the gate offset $n_g$, which have been corrected in the cuts shown in panels (f) and (i). 
}
\end{figure*}

\begin{figure}
\includegraphics{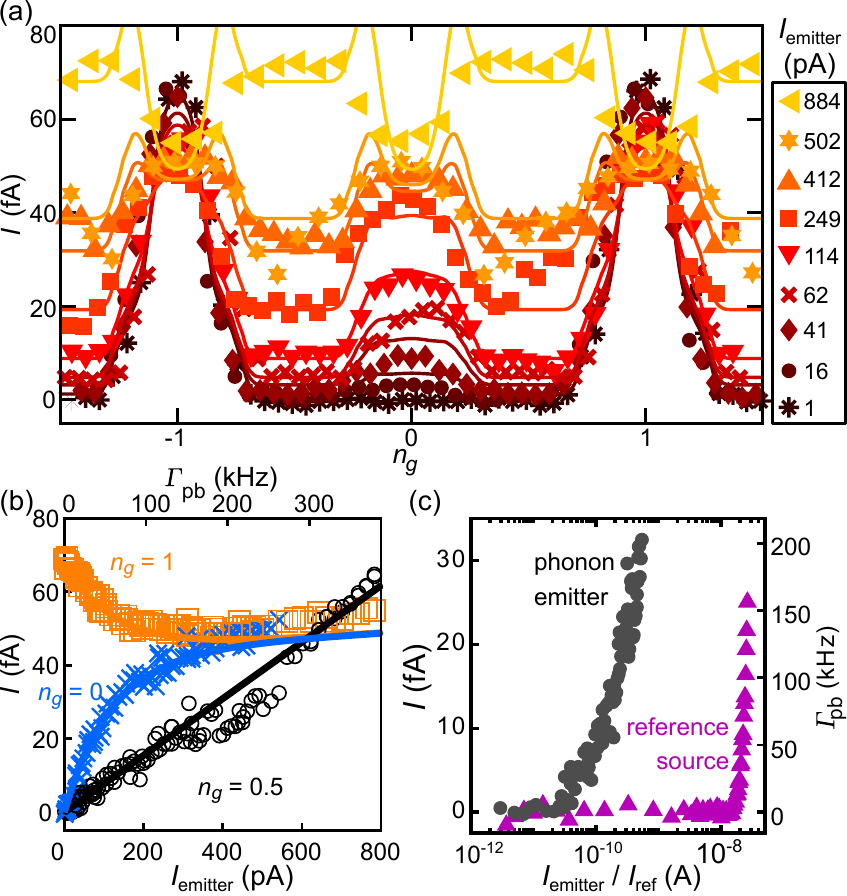}
\caption{\label{fig:gammapb}
(a) Detector current versus gate offset $n_g$ at $V_{b}=$ 110 $\mu$V with varying $I_{\text{emitter}}$ in the experiment (symbols) and different Cooper pair breaking rates $\Gamma_{\text{pb}} = A I_{\text{emitter}}/e$ in the simulations (solid lines). $A=7.8\times 10^{-5}$ is used for all simulated curves. 
(b) $I$ extracted at the self-calibrating operating point ($n_g=0.5$, circles) and for comparison at $n_g=1$ and $n_g=0$, where the current is due to Andreev reflection. Simulated curves (solid lines) for all three values of $n_g$ are calculated assuming $\Gamma_{\text{pb}} = A I_{\text{emitter}}/e$ with $A=7.8 \times 10^{-5}$.
(c) Measured $I$ at $n_g=0.5$ and the corresponding pair-breaking rate in the detector, versus current through either the phonon emitter (black circles) or reference source (magenta triangles).
}
\end{figure}

A scanning electron micrograph of our proof-of-concept device is shown in Fig. 1(d). The pair-breaking detector is an aluminum island with $\mathcal{V}=0.9 \times 0.4 \times 0.08~\mu$m$^3$, $\Delta=200~\mu$eV and $E_C=92~\mu$eV, tunnel coupled to normal metallic copper leads. We fabricate a superconducting phonon emitter and normal metallic reference source on a conducting silicon substrate simultaneously with the detector. The conducting substrate is not necessary for detector operation. Measurements were performed in a plastic dilution refrigerator at a base temperature of 40 mK. In measurements with the aluminum in the normal state, we extract device parameters, including the electron-phonon coupling constant $\Sigma_{\text{Al}} \approx 2 \times 10^8$ W K$^{-5}$m$^{-3}$, and verify that heat conduction through the substrate by thermal phonons is negligible as long as the dissipated power is below 1 pW \cite{Note1}.

Figure \ref{fig:pinkabsorber} presents the operation of the self-calibrating detector in the superconducting state. 
When the current through both the phonon emitter and the reference source is zero, the current at $n_g = 0.5$ and low $V_b$ is zero within the measurement accuracy of roughly 1 fA [Fig. \ref{fig:pinkabsorber}(a-c)] as expected. At around $n_g=1$, the current is finite due to Andreev reflection \cite{hekking1993coulomb, eiles1993evenodd}. When we increase the current through the phonon emitter to 30 pA [Figs. \ref{fig:pinkabsorber}(d-f)] or 120 pA  [Figs. \ref{fig:pinkabsorber}(g-i)], the current level at the plateau  around $n_g = 0.5$ and $|V_b| \approx 100~\mu$V increases, as predicted by our simple model. The current increases also around $n_g=0$, which is due to Andreev current flowing once the odd charge states are populated due to quasiparticles \cite{hergenrother1994charge, hergenrother1995photonactivated}.

In contrast, when we increase the current through the reference source up to 7.5~nA,  while the phonon emitter is kept grounded [Fig. \ref{fig:pinkabsorber}(j-l)], the measured current stays zero within the $2e$-periodic Coulomb diamonds. The main difference between the phonon emitter and the reference source is that the emitter island is superconducting aluminum, which will emit phonons whose energy distribution is peaked above $2\Delta \approx k_B \times$~4.6~K when overheated even slightly \cite{eisenmenger1967quantum}. The reference island is normal metallic copper, which will instead emit a broad thermal distribution of phonons, and the superconducting leads of the reference source are connected to their normal metal shadow copies acting as quasiparticle traps. Hence the reference source will emit orders of magnitude fewer phonons with energy larger than $2\Delta$ than the phonon emitter. \nocite{hergenrother1995photonactivated, naaman2007narrowband, jalalijafari2016detection}. This proves that the measured response is indeed due to pair-breaking radiation, unlike in other detector proposals utilizing similar devices \cite{hergenrother1995photonactivated, naaman2007narrowband, jalalijafari2016detection}.

Our experimental data are quantitatively reproduced by numerical simulations, shown in Figs. \ref{fig:pinkabsorber} and \ref{fig:gammapb}. 
We obtain excellent agreement by changing only $\Gamma_{\text{pb}}$ for differing $I_{\text{emitter}}$. In Fig. \ref{fig:gammapb}(a) we show cuts in the data at $V_b = 110~\mu$eV with differing $I_\text{emitter}$. At all but the lowest emitter currents, the data agrees well with simulations assuming a linear relation $\Gamma_{\text{pb}} = A I_{\text{emitter}}/e$ with $A=7.8 \times 10^{-5}$. Thus our simulation accounts for all the relevant processes, which verifies that the self-calibrating mode can be used in experiments. The points in Fig. \ref{fig:gammapb}(b) and \ref{fig:pinkabsorber}(c) where the measured current is lower than expected, and the poorer agreement of the data with the simulations at $I_{\text{emitter}}=502$~pA in Fig. \ref{fig:gammapb}(b), are most likely due to an external disturbance during part of the sweep \cite{Note1}.

% In Fig. \ref{fig:gammapb}(a) we show ... xx

Next, we turn to measure the phonon transmission from the phonon emitter to the detector. Modeling of the recombination phonon emission \cite{Note1} allows us to estimate the emission rate of phonons with energy $\geq 2\Delta$ as  $\Gamma_{2\Delta} = \eta I_{\text{emitter}}/e$ with the proportionality constant $\eta=0.68$. 
We find that the measured detector current is also linear in $I_{\text{emitter}}$ at $n_g = 0.5$, as shown in Fig. \ref{fig:gammapb}(b). Hence we extract the proportion of phonons emitted that are absorbed in the detector as $x  = \Gamma_{\text{pb}}/\Gamma_{2\Delta} \approx 1.1 \times 10^{-4}$. This fraction is difficult to estimate by other means due to the long mean free paths of the phonons at low temperatures, but the order of magnitude is in line with measurements in Refs. \cite{otelaja2013design, hertzberg2014direct}. We also quantify the difference in Cooper pair breaking caused by the phonon emitter and reference source in Fig. \ref{fig:gammapb}(c). The reference source causes no detectable signal until currents of almost 20 nA are applied, even when we obtain a measurable signal from as little as 16 pA passing through the superconducting phonon emitter, which corresponds to less than 10~fW of dissipated power.

As a detector of phonons with energy $2\Delta$, our device has a noise equivalent power of $3 \times 10^{-18}$~W/$\sqrt{\text{Hz}}$.  In calculating this figure, we use the measured noise level and assume that 25\% of incident phonons are absorbed in the detector \cite{Note1}. This is over two orders of magnitude better than in the microscale phonon detectors of Ref. \cite{otelaja2013design}, while our device also provides the advantage of a self-calibrating operating mode and well-defined absorption volume. 
The minimum detectable signal in an 1 s integration time is $\Gamma_\text{pb}=1$~kHz. The maximum output signal of the detector is limited by Eq. \eqref{eq:recombinationlimit}, yielding a dynamic range that covers several orders of magnitude.
The device performance could be enhanced at other operating points once the response has been calibrated in the self-calibrating mode. In our device, the responsivity at low phonon flux increases by roughly a factor of 2 at $n_g = 0$ (Fig. \ref{fig:gammapb}(b)),  
and could be increased further at higher $V_b$.

\paragraph{Conclusions.}

In conclusion, we have proposed and experimentally implemented a mesoscopic superconducting detector of high-frequency radiation leading to Cooper pair depairing. We have used the proof-of-concept device to detect the nonequilibrium phonons emitted by another superconducting detector, while using a reference source to rule out mechanisms other than pair breaking. 
Due to the well-defined microscale absorption volume and self-calibrating operation, our detector could be particularly useful for studying athermal phonon propagation, relevant both in the context of low-temperature detectors \cite{martinez2019measurements, karatsu2019mitigation, henriques2019phonon} and in mitigating phonon-mediated quasiparticle poisoning of superconducting quantum circuits \cite{patel2017phononmediated, leonard2019digital, wilen2020correlated}. 
As a first step in this direction, we extract the fraction of phonons transmitted from emitter to detector over a distance of 8 $\mu$m. As the detector is easily adaptable to different geometries, our device would be straightforward to integrate to study e.g. the phonon transmission across a typical qubit chip, or fabricate on the sidewalls of a mesa structure for phonon spectroscopy applications  \cite{otelaja2013design}.
As low a power as 10 fW dissipated in the phonon emitter caused a measurable increase in the Cooper pair creation rate on the absorber island. Hence phonon-mediated poisoning is a plausible explanation of the charge detector backaction of Refs. \cite{mannik2004effect, mannila2019detecting}, and our results highlight the importance of avoiding dissipation in superconducting quantum devices.  
Finally, we expect that monitoring the individual quasiparticle relaxation events with a fast charge detector \cite{qp-statistics} would enable detecting every single phonon or photon absorbed.

\begin{acknowledgments}
We acknowledge useful discussions with O. Maillet and J. T. Peltonen. 
This work was performed as  part of the Academy of Finland Centre of Excellence program (project 312057).
We acknowledge the provision of facilities and technical support by Aalto University at OtaNano - Micronova Nanofabrication Centre and OtaNano - Low Temperature Laboratory.
E.T.M. and J.P.P. acknowledge financial support from Microsoft. 
V.F.M. acknowledges financial support from the Swedish National Science Foundation, the QuantERA project “2D hybrid materials as a platform for topological quantum computing”, and NanoLund.
\end{acknowledgments}

%\bibliography{../phonons}


\end{document}

\title{A self-calibrating superconducting pair-breaking detector \\ Supplemental Material}

\author{E. T. Mannila}
\email{elsa.mannila@aalto.fi}
\affiliation{QTF Centre of Excellence, Department of Applied Physics, Aalto University, FI-00076 Aalto, Finland}

\author{V. F. Maisi}
\affiliation{Physics Department and NanoLund, Lund University, Box 118, 22100 Lund, Sweden}

\author{J. P. Pekola}
\affiliation{QTF Centre of Excellence, Department of Applied Physics, Aalto University, FI-00076 Aalto, Finland}

\date{\today}

\renewcommand{\thefigure}{S\arabic{figure}} 
\renewcommand*{\citenumfont}[1]{S#1}
\renewcommand*{\bibnumfmt}[1]{[S#1]}

\begin{abstract}
This supplementary information includes details on sample fabrication and normal-state characterization, and modeling of phonon emission, and calculations for the limits on the operation of the self-calibrating mode.  
\end{abstract}

\maketitle

\section{Device fabrication and normal-state characterization}

\subsection{Fabrication}

The three single-electron transistors in our experiment, shown in Fig. 1(d) of the main text, are fabricated on the same silicon chip approximately 8 $\mu$m from each other. The substrate is a 525 $\mu$m thick silicon (100) boron-doped substrate (resistivity 0.001-0.005 $\Omega$cm) that stays conducting at low temperatures, coated with 200 nm of thermal silicon oxide. The conducting substrate was chosen in order to avoid fabricating separate ground planes to suppress photon-assisted tunneling events \cite{pekola2010environment} in our measurement setup, which is not fully microwave tight. 
Similar substrates are commonly used in experiments on semiconducting nanowires, and the same substrate has been used as a back gate in e.g. Refs. \cite{taupin2016inas, higginbotham2015parity}. However, in a well-shielded measurement setup, like in Ref. \cite{hergenrother1995photonactivated}, ground planes or a conducting substrate would not be needed. 

Due to the insulating oxide, the distance between the SETs, and the conducting substrate, there is no direct resistive or capacitive coupling between the islands of the SETs, although there is some capacitive crosstalk between the bulky gate electrodes.
The devices are fabricated in one deposition step with standard electron-beam lithography and two-angle evaporation. A thin aluminum oxide tunnel barrier separates the 80 nm superconducting aluminum and 100 nm normal metal copper films, which form the island and leads, respectively, of the detector and phonon emitter. In the reference source, the island is copper and leads aluminum, but the design is otherwise as similar as possible to that of the phonon emitter. The phonon emitter island has a relatively large volume $\mathcal{V}$ to obtain a large phonon emission rate with a given temperature. 
The normal metal leads of the detector are relatively bulky and extend to the bonding pads, but we foresee that they could be replaced by small normal metal sections connected to superconducting leads and bonding pads. A normal metal section a few micrometers long should be enough to suppress the proximity effect due to the superconducting leads \cite{karimi2020optimized}, and thus we believe that our device would be straightforward to adapt to different geometries.% integrate to study e.g. the phonon transmission across a typical qubit chip, or fabricate on the sidewalls of a mesa structure for phonon spectroscopy applications like in Ref. \cite{otelaja2013design}.

\subsection{Measurements}

The silicon chip is attached with vacuum grease to a brass sample holder thermally anchored to the mixing chamber of a home-made plastic dilution refrigerator with a base temperature of 40 mK. The sample holder is closed with a single threaded cap, which together with the ground plane formed by the conducting substrate suppresses photon-assisted tunneling due to stray microwaves \cite{hergenrother1995photonactivated, pekola2010environment} adequately for this experiment, although the sample stage is possibly not fully microwave-tight. 
The measurement wiring lines are filtered with approximately 2 m of Thermocoax \cite{zorin1995thermocoax} between the 1 K and mixing chamber stages, which limits the achievable bandwidth to the kHz range. Direct current electrical measurements were performed with room-temperature voltage sources and transimpedance amplifiers. For the subgap measurements of the detector, the amplifier used was a Femto model LCA-2-10T set to 2~Hz or 0.3~Hz bandwidth and $10^{12}$~V/A gain. 

In the subgap measurements shown in Figs. 2 and 3 in the main text, we have subtracted a leakage current linear in the gate voltage corresponding to a resistance of 10$^{12}$ $\Omega$, most likely originating from the room-temperature breakout box. The measurements shown in Figs. 2(a,d,g,j) in the main text were performed by sweeping the gate voltage tuning $n_g$ at successive values of the bias voltage. The gate offset $n_g$ drifts and occasionally jumps over timescales of minutes or hours. This contributes to the apparent skew of the experimental panels of Fig. 2, and several abrupt jumps of $n_g$ are also visible  in panels 2(d) and 2(g). The jumps have been corrected for the cuts shown in Fig. 2(f) and 2(h). The rate of the jumps or drifts of $n_g$ had no strong dependence on the value of gate or bias voltages applied. Part of the skew can be also due to the asymmetrical biasing scheme used or a slight asymmetry of the junction capacitances, both of which cause $n_g$ to depend on the bias voltage $V_b$ in addition to the gate voltage.

The data shown in Figs. 1(c) and 3 of the main text are measured by sweeping $n_g$ at successive values of $I_\text{emitter}$ with the emitter either current or voltage biased, and several such datasets are combined in the figures. In part of one of the datasets, corresponding to the points around $\Gamma_{\text{pb}}=200~$kHz lying under the simulated and ideal lines in Figs. 1(c) and 3(b), the data does not form clear plateaus as a function of $n_g$, although the response at $n_g = 0.5$ is quite close to the ideal and simulated values. The data in Fig. 3(a) of the main text corresponding to $I_\text{emitter}=502$~pA are part of this dataset. In this case, the detector response is sensitive to the precise gate voltage value, which may contribute to the discrepancy between the ideal and measured current. We believe that this is due to an external influence on the measurement during that particular sweep and is not a characteristic of the detector itself, as clear plateaus are visible in other datasets aqcuired at similar or higher detector currents, such as in the sweeps at $I_\text{emitter} = 412$~pA and $I_\text{emitter} = 884$~pA in Fig. 3(a). 

\subsection{Normal-state characterization}

\begin{figure}
\includegraphics{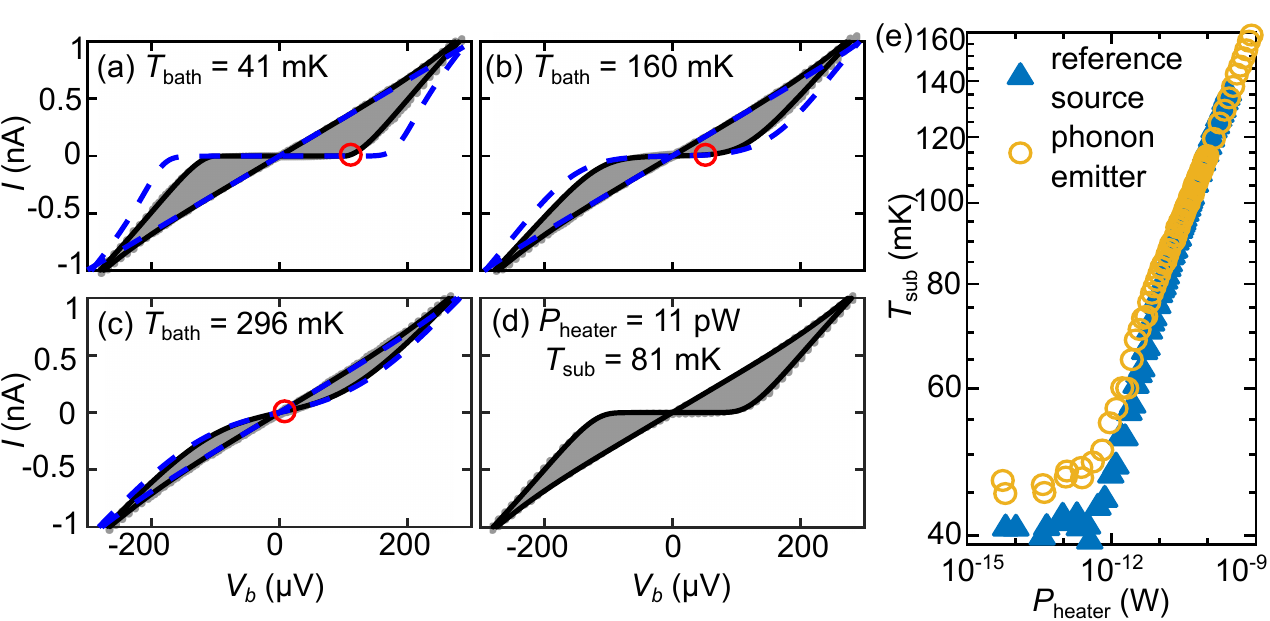}
\caption{\label{fig:normal}(a-c) Measured and simulated $I$-$V_b$ characteristics of the detector, measured in the normal state with a magnetic field $B_\perp$ = 70 mT applied parallel to the sample at different bath temperatures $T_{\text{bath}}$. Gray dots show the measured current when the gate voltage is swept over several periods of the gate offset $n_g$, while black lines show simulations with $T_{\text{is}}$ solved self-consistently at $n_g$=0 (minimum current) and $n_g$=0.5 (maximum current) with parameters given in Table \ref{table1}. Dashed lines show simulations where the overheating of the islands is neglected and $T_{\text{is}}=T_{\text{bath}}$. Red circles show the maximum voltage at a current of 10 pA, used as a thermometer. 
(d) Detector I-V characteristics measured at base temperature but with 11 pW heating power applied to the reference source, superimposed with simulations assuming a bath temperature of 81 mK. 
(e) Extracted substrate temperature when heater power $P_{\text{heater}}$ is applied to the phonon emitter (circles) or reference source (triangles) in the normal state at $T_{\text{bath}} = 40~$mK. 
}
\end{figure}

We characterize the devices with the aluminum driven into the normal state by a magnetic field of 70 mT applied perpendicular to the substrate. 
For each device, we simultaneously fit current-voltage characteristics measured at bath temperatures $T_{\text{bath}}$ between 41 mK and 441 mK (Fig. \ref{fig:normal}(a-c)) with a standard rate equation model for a normal-metallic single-electron transistor. In the simulations, we allow the island temperature $T_{\text{is}}$ to deviate from $T_{\text{bath}}$ and solve $T_{\text{is}}$ self-consistently from the heat balance equation  \cite{giazotto2006opportunities}
\begin{equation}\label{eq:heatbalance}
\dot{Q}_\text{tunn} = \dot{Q}_\text{e-ph},
\end{equation} where $\dot{Q}_\text{tunn}$ is the heat current due to the tunneling electrons. The heat flux to the phonon bath is given by 
\begin{equation}\label{eq:ephnormal}
\dot{Q}_\text{e-ph} = \Sigma \mathcal{V} (T_{\text{is}}^5-T_{\text{bath}}^5),
\end{equation}
where $\Sigma$ is the electron-phonon coupling constant and $\mathcal{V}$ the volume. From these fits, we determine the total tunnel resistance $R_T$ and charging energy $E_C=e^2/2C_\Sigma$, where $C_\Sigma$ is the total capacitance of the island, given in Table \ref{table1}.  The fits shown in Fig. \ref{fig:normal} are obtained with $\Sigma_{\text{Al}} = 2 \times 10^8$~WK$^{-5}$m$^{-3}$, consistent with results in the literature \cite{kautz1993selfheating, kauppinen1996electronphonon, meschke2004electron}. However, our fit would be consistent with values $\Sigma_{\text{Al}}$ between $1 \times 10^8$ to $10 \times 10^8$ WK$^{-5}$m$^{-3}$ and also power-law exponents in Eq. \ref{eq:ephnormal} differing from 5.
We determine $\Delta$ from similar measurements in the superconducting state at zero magnetic field. For the copper island of the reference source, we use the literature value $\Sigma_{\text{Cu}} = 2 \times 10^9$ WK$^{-5}$m$^{-3}$ \cite{giazotto2006opportunities, wang2019crossover}. 
  
In the numerical simulations used in the superconducting state shown in Figs. 2 and 3 in the main text, we use instead a rate equation tracking the probabilities of having $N$ excess electrons and $N_{\text{QP}}$ quasiparticles on the island, with the rates given in Ref. \cite{maisi2013excitation}. In addition to the parameters determined from large-scale IV fits, we fit the effective conduction channel area $A_{ch} \approx 15$~nm$^2$ setting the magnitude of the Andreev current from the slope of the current at $n_g = 1$, consistent with the values of Ref. \cite{maisi2011realtime}. 

\begin{table*}
\caption{\label{table1}
Device parameters. The tunnel resistance $R_T$, charging energy $E_C$ and superconducting gap $\Delta$ were determined by fitting large-scale IV curves measured in the normal ($R_T$, $E_C$) or superconducting state ($\Delta$) as described in the text. Junction asymmetry is defined as the ratio between the areas and hence resistances of the two junctions in each device. The island and junction areas were estimated from micrographs of the devices. 
}
\begin{ruledtabular}
\begin{tabular}{l|lllllll}
Device & Island material & $\Sigma$ ($10^8$ WK$^{-5}$m$^{-3}$) & $\mathcal{V}$ ($\mu$m$^3$) & $\Delta$ ($\mu$eV) & $R_T$ (k$\Omega$) & $E_C$ ($\mu$eV) & Junction asymmetry \\ \hline
Detector & Al & 2 & 0.9 $\times$ 0.4 $\times$ 0.08 & 200 & 151 & 92 & 1\\
Phonon emitter & Al & 2 & 1.8 $\times$ 0.8 $\times$ 0.08 & 200 & 87 & 70 & 1 \\
Reference source & Cu & 20 & 1.8 $\times$ 0.7 $\times$ 0.1 & 195 & 21 & 20 & 3\\
\end{tabular}
\end{ruledtabular}
\end{table*}

We characterize the heat conduction of the substrate due to thermal phonons with the devices in the normal state. The current through a normal-metallic SET is sensitive to the temperature of the metals, and effectively we can use one of the SETs as a Coulomb blockade thermometer \cite{pekola1994thermometry}, although in the unusual regime $E_C > k_B T$. As a probe of the substrate temperature $T_{\text{sub}}$, we use the maximum voltage drop over the detector at fixed bias current of 10 pA, indicated by red circles in Figs. \ref{fig:normal}(a-c). The voltage drop is measured as a function of $T_{\text{bath}}$ to obtain a calibration. 
We then repeat the measurement at $T_{\text{bath}}\approx$ 40 mK but using either the emitter or reference source as a heater dissipating a measured power $P_{\text{heater}} = IV$. We have verified by comparing full IV curves measured at finite $P_{\text{heater}}$ with simulations to verify that in the normal state, the only effect of the heaters is to increase the substrate temperature $T_{\text{substrate}}$ seen by the detector (Fig. \ref{fig:normal}(d)).
We observe no change in $T_{\text{sub}}$ below $P_{\text{heater}} \approx$ 1 pW, see Fig. \ref{fig:normal}(e), and at higher powers the dependence follows approximately a power law. As expected for thermal conduction through the substrate, the response to heater power is similar when applied to the phonon emitter or the reference source, and the order of magnitude is similar to that observed in earlier reports \cite{savin2006thermal, feshchenko2017thermal}. 

\section{Modeling of phonon emission}

\begin{figure}
\centering
\includegraphics{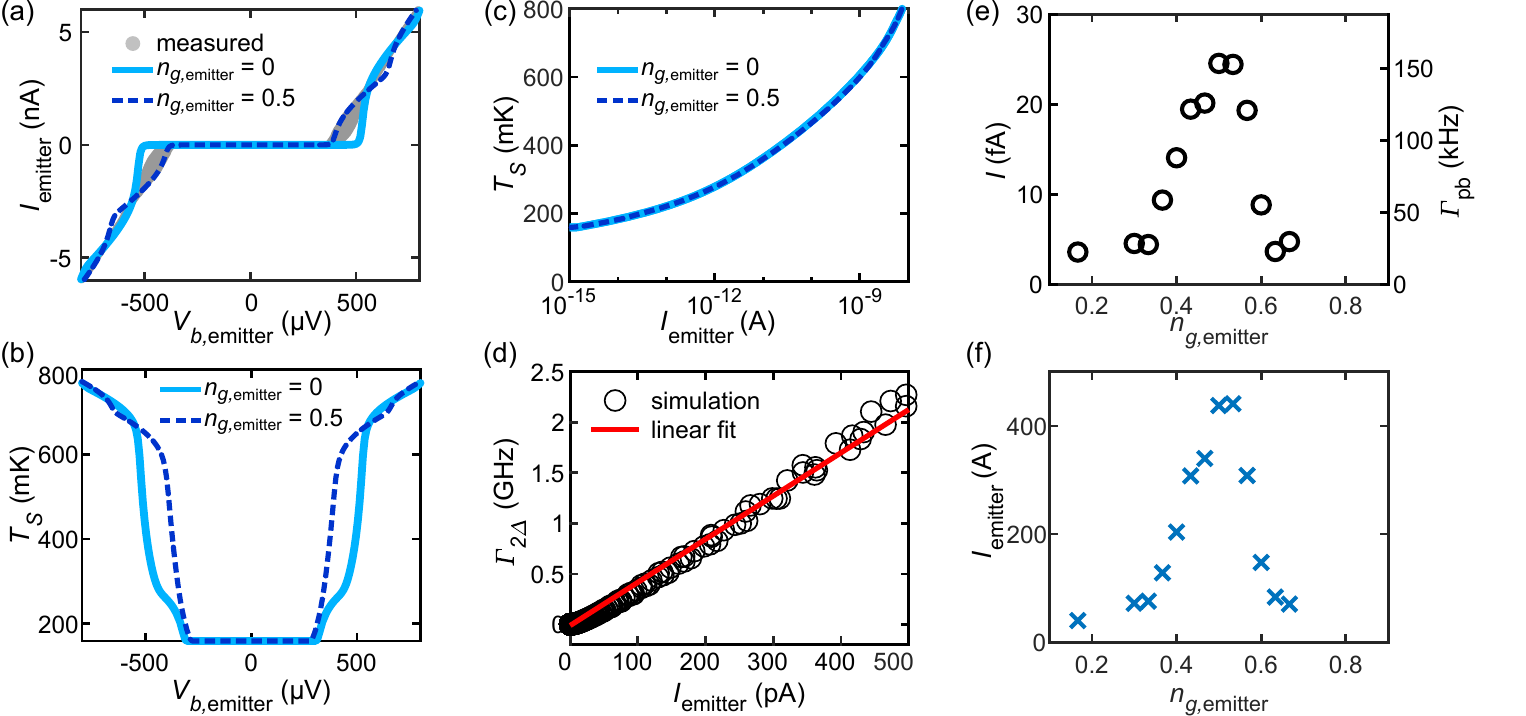}
\caption{\label{fig:emitter} (a) Current-voltage characteristics of the phonon emitter in the superconducting state. Measurements (gray) are swept over several periods of $n_{g,\text{emitter}}$, while the simulations show the minimum and maximum current at each $V_{b,\text{emitter}}$, obtained at $n_{g,\text{emitter}}=0$ (solid lines) and $n_{g,\text{emitter}}=0.5$ (dashed lines), respectively.
(b) The self-consistently solved temperature $T_S$ on the superconducting emitter island versus voltage.
(c) The emitter temperature as a function of $I_{\text{emitter}}$. The temperature for all values of $n_{g,\text{emitter}}$ collapse on top of each other, and the emitter starts to overheat already at currents below 1 pA. 
(d) Rate of recombination phonon emission $\Gamma_{2\Delta}$ according to Eq. \ref{eq:recrate}, calculated from the simulated $T_S$ shown in panels (b) and (c) (circles), and a linear fit between $\Gamma_{2\Delta}$ and $I_{\text{emitter}}$ (solid line).
(e) Detector current $I$ and the corresponding Cooper pair breaking rate $\Gamma_{\text{pb}}$, measured at fixed emitter voltage bias $V_{b,\text{emitter}}$ = 395 $\mu$V but varying emitter gate offset $n_{g,\text{emitter}}$. 
(f) Measured $I_{\text{emitter}}$ corresponding to the data shown in panel (e).
}
\end{figure}

To obtain the emission rate of phonons from a superconductor, one could solve the kinetic equations for the nonequilibrium quasiparticle and phonon distributions within the superconductor with the voltage bias serving as a quasiparticle injection term \cite{chang1977kineticequation, patel2017phononmediated}. However, here we use a simpler model assuming a thermal distribution of quasiparticles at an elevated temperature $T_S$, coupled to a thermal phonon distribution at $T_{\text{bath}}$. Although the distribution of the quasiparticles may not follow a thermal form, we believe a thermal model adequate here for the following reasons: 
 First, the bias voltage across the two tunnel junctions of our emitter satisfies $V_{b,\text{emitter}}<2.3 \Delta/e$, so that the relaxation phonons emitted by the injected quasiparticles will not have enough energy to break further Cooper pairs.
Second, the current in a normal metal-insulator-superconductor structure biased above the gap is rather insensitive to the number or distribution of quasiparticles. 
Third, the recombination rate itself does not depend strongly on the distribution of quasiparticles but rather on their total number. Thus we use $T_S$ as a convenient quantity for parametrizing the number of quasiparticles. 
The effects of phonon trapping \cite{rothwarf1967measurement} can be included in the value of the electron-phonon coupling constant $\Sigma$, as discussed below.

We obtain $T_S$ by solving the heat balance equation, Eq. \eqref{eq:heatbalance}, but with both the tunneling rates and electron-phonon coupling modified by superconductivity. 
The rates and heat deposited in single-electron tunneling are given for instance in \cite{giazotto2006opportunities}. For the electron-phonon heat-flow we use Eq. (3) from Ref. \cite{maisi2013excitation}.
We solve self-consistently the emitter current $I_{\text{emitter}}$, temperature $T_S$ and superconducting gap $\Delta(T_S)$ of the island as a function of $n_{g,\text{emitter}}$ and $V_{b,\text{emitter}}$, as shown in Fig. \ref{fig:emitter}. In the regime $I_{\text{emitter}}<1$ nA, relevant for the experiment, $\Delta(T_S)$ is suppressed from the zero-temperature value by less than 3\%.%, although at the highest emitter currents this gap suppression might already influence the energy distribution of the emitted phonons. 
We also include a small parasitic heat flow $P_0=10^{-21}$ W into the island to improve numerical convergence, which does not affect the results when $T_S>200$ mK, the range relevant for phonon emission.
After solving $T_S$, we then calculate the emission rate of phonons with energy $\geq2\Delta$ from the recombination rate \cite{maisi2013excitation}
\begin{equation}\label{eq:recrate}
\Gamma_{2\Delta} = \frac{\dot{Q}_{R}}{2\Delta} = \frac{\pi \mathcal{V} \Sigma}{6 \zeta(5) k_B^5} \left[ k_B T_S \Delta^3 + \frac{7}{4}(k_B T_S)^2 \Delta^2 \right] e^{-\frac{2\Delta}{k_B T_S}},
\end{equation}
where $\zeta$ is the Riemann zeta function. 

Both the current $I_{\text{emitter}}$ and island temperature $T_S$ depend on the gate offset $n_{g,\text{emitter}} = V_{g,\text{emitter}} C_{g,\text{emitter}}/e$ when calculated at fixed $V_{b,\text{emitter}}$, as shown in Fig. \ref{fig:emitter}(a-b). However, for a fixed $I_{\text{emitter}}$, the simulated $T_S$ does not depend on $n_{g,\text{emitter}}$, but instead all the curves collapse on each other [Fig. \ref{fig:emitter}(c)]. The island is also overheated very rapidly due to the exponential suppression of the electron-phonon coupling: a current of 1 pA is enough to heat the emitter up to 270 mK. The 2$\Delta$ phonon emission rate in Fig. \ref{fig:emitter}(d) calculated with Eq. \ref{eq:recrate} is close to linear in $I_{\text{emitter}}$. We fit a linear dependence between the $2\Delta$ phonon emission rate and emitter current [solid lines in Fig. \ref{fig:emitter}(d)], yielding $\Gamma_{2\Delta} = \eta I_{\text{emitter}}/e$ with $\eta = 0.68$. 
The simulations shown in Figs. 2 and 3 of the main text correspond to setting $\Gamma_{\text{pb}} = A I_{\text{emitter}}/e$, where the proportionality constant $A=7.8 \times 10^{-5}$.
This allows us to deduce $\Gamma_{\text{pb}} = x \Gamma_{2\Delta}$, where $x = A/\eta = 1.1 \times 10^{-4}$ is the fraction of $2\Delta$ phonons emitted that are absorbed in the detector. 

At the smallest currents ($I_{\text{emitter}}<50$ pA), the measured detector current shown in Fig. 3(a) of the main text is somewhat smaller than the simulation assuming the linear relation between $I_{\text{emitter}}$ and $\Gamma_{2\Delta}$. This could be explained by a relatively large part of the current at the smallest $I_{\text{emitter}}$ being carried by Andreev reflection, which does not heat the superconductor and thus causes no phonon emission. Also, the linear fit in Fig. \ref{fig:emitter}(d) overestimates the phonon emission somewhat compared to the simulated $\Gamma_{2\Delta}$. We note that if the emitter is biased to subgap voltages $V_{b,\text{emitter}} \leq 2 \Delta$, the single-electron current through the emitter is strongly suppressed and the phonon emission rate is zero, as well. As expected, in the experiment there was no change in the detector response when either of the emitters was biased to subgap voltages ($V_{b,\text{emitter}} < 2\Delta/e$), in contrast to Ref. \cite{naaman2007narrowband}, where quasiparticle tunneling rates were measured as a function of the voltage bias of a fully superconducting Cooper pair transistor expected to emit Josephson-like radiation at subgap voltages. 

Our simulations in Fig. \ref{fig:emitter}(a) do not reproduce the rapid rise of $I_{\text{emitter}}$ at $n_{g,\text{emitter}} = 0$ and $V_{b,\text{emitter}}$ slightly above $2\Delta/e$. We attribute this to inelastic cotunneling \cite{averin1990virtual, averin1992singleelectron, eiles1993evenodd}, which is not included in our model, but argue that this does not substantially change the rate of $2\Delta$ phonon emission or the functional dependence on current. In an inelastic cotunneling event, two electrons simultaneously tunnel through the whole device, leaving two excitations on the island. However, in two successive single-electron tunneling events, two excitations on the island are involved as well. Because $V_{b,\text{emitter}}$ is close to $2\Delta/e$, the energies of all the excitations will be close to $\Delta$, and we expect that inelastic cotunneling will lead to a similar dependence of $T_S$ and hence $\Gamma_{2\Delta}$ on $I_{\text{emitter}}$ as single-electron tunneling. 

We can also estimate the nonequilibrium phonon emission from the superconducting leads of the reference source by considering quasiparticle diffusion in the spirit of Ref. \cite{knowles2012probing}. We assume a simplified geometry with superconducting leads with a rectangular cross section with area 80 nm $\times$ 200 nm, coupled to a perfect quasiparticle trap at a distance of $1~\mu$m. In the experiment, the lead widens much faster and overlaps with the imperfect quasiparticle trap formed by the normal shadow already at half a micron distances, so this leads to an upper limit for the quasiparticle temperature at the tunnel junction. We use $\rho =$ 90 $\Omega$nm as the normal-state resistivity of the aluminum film \cite{marinsuarez2020active}. These values lead to 330 mK temperature at the junction with $I_{ref} = 10$~nA. The volume 80 nm $\times$ 200 nm $\times$ 1 $\mu$m at such a temperature 
would emit recombination phonons at a rate of 500 kHz, two orders of magnitude lower than the emission rate from the phonon emitter at $I_{\text{emitter}}=15~$pA.
However, recombination phonons emitted by the leads could be a plausible explanation for the current-dependent quasiparticle poisoning due to the backaction of a charge detector with much longer isolated superconducting leads and a normal island in Ref. \cite{mannila2019detecting}, as well as explaining the backaction from the fully superconducting devices in Ref. \cite{mannik2004effect}. 

\section{Estimation of phonon trapping effects and phonon absorption probability}

In this section, we estimate the effects of phonon trapping (phonon reabsorption before it escapes to the substrate) in the phonon emitter, as well as the fraction of incident phonons that are absorbed in the detector. The estimates have relatively large uncertainties, but we note that due to the self-calibrating operation our device could be useful in measuring the phonon transmission and absorption probabilities if it were subjected to a known phonon flux. 

In the phonon emitter, as the timescales for phonon absorption (Cooper pair breaking) are several orders of magnitude faster than the recombination times \cite{kaplan1976quasiparticle}, the effect of phonon trapping is to simply increase the recombination times by some material- and substrate-dependent factor $F$, which we can view as decreasing the electron-phonon coupling constant $\Sigma$ by the same factor. We can estimate the phonon trapping factor as $F=4 d/ \eta \Lambda$ \cite{kaplan1979acoustic}, where $d=80$~nm is the film thickness, $\eta \approx 0.7$ is the transmission probability from the film to the substrate \cite{kaplan1979acoustic, oneil2012measurement}, and $\Lambda$ is the mean free path for phonons of energy $2\Delta$ against pair breaking, for which Ref. \cite{kaplan1979acoustic} cites values between 100 nm and 350 nm. With these values, the phonon trapping factor ranges from 1.3 to 4, while using the somewhat longer mean free paths measured for 100 GHz phonons measured in the normal state in \cite{mrzyglod1994mean} corrected to the superconducting state \cite{bobetic1964evaluation}, $F$ is even closer to unity. As our value for $\Sigma$ measured in the normal state is somewhat on the low side compared to the literature values, we do not further correct for phonon trapping. 

For the phonon absorption probability in the detector, we may consider that Ref. \cite{martinez2019measurements} found good agreement between measurements and simulations using the acoustic mismatch model with a transmission coefficient $T_{Si \rightarrow Al}$ between 0.3 and 0.55 for 60 nm Al films, which assumed that the phonon is absorbed with an unity probability once it entered the film. In contrast, Ref. \cite{otelaja2013design} assumed a transmission probability from Si to Al larger than 0.9 but phonon absorption probabilities in 140-220 nm thick films on the order of 0.25. In contrast to these works, in our case the Si substrate is covered by a 200 nm SiO$_2$ film, and to our knowledge there are no experimental data available for such a system. As the acoustic impedances are better matched between Al and SiO$_2$ than Al and Si \cite{kaplan1979acoustic}, we believe 25\% to be a reasonably cautious estimate for the fraction of incident phonons that are eventually absorbed in the Al film. This parameter is only used in the calculation of the noise-equivalent power.

\section{Additional derivations}

In this section, we show that the self-calibrating response is obtained even if the tunneling rates through the two junctions are unequal, derive the condition of Eq. (3) given in the main text, as well as the range of $V_b$ and $n_g$ for the self-calibrating mode. 

\subsection{Unequal tunnel resistances}

Here, we consider a system where the tunneling rates for quasiparticles leaving the island differ, which in practice may be due to different junction areas and resistances. Instead of the single rate $\Gamma_{\text{tunn}}$, the quasiparticles may exit the island through the left and right junctions with rates $\Gamma_\text{L}$ and $\Gamma_\text{R}$, respectively, but we still assume that the return from the charge states $N=-1$ or +2 happens instantaneously compared to either $\Gamma_\text{L}$ or $\Gamma_\text{R}$. Now, we can write a rate equation for the probability $P_{N_\text{QP}}$ to have $N_\text{QP}$ quasiparticles present: 
\begin{equation}\label{eq:selfcalibrating-rate-for-nqp}
\frac{d}{dt} P_{N_\text{QP}} = -\Gamma_{\text{pb}} P_{N_\text{QP}} + \Gamma_{\text{pb}} P_{N_\text{QP}-2} - (\Gamma_\text{L} + \Gamma_\text{R}) N_\text{QP} P_{N_\text{QP}} + (\Gamma_\text{L} + \Gamma_\text{R}) (N_\text{QP}+1) P_{N_\text{QP}+1} 
\end{equation}
The probabilities $P_{N_\text{QP}}$ for negative $N_\text{QP}$ are set to zero, and in the steady state the time derivative is zero. Multiplying by $N_\text{QP}$ and summing over all $N_\text{QP}$ we obtain the steady state quasiparticle population
\begin{equation}
\langle N_\text{QP} \rangle = 2 \Gamma_{\text{pb}} / (\Gamma_\text{L} + \Gamma_\text{R}).
\end{equation}
To calculate the total current, we need the rate at which the first step in the cycle $N=0 \rightarrow -1 \rightarrow 0$ ($N=+1 \rightarrow +2 \rightarrow +1$) occurs. The first step is a quasiparticle tunneling out through the left (right) junction. From a state with $N_\text{QP}$ quasiparticles, this occurs at a rate $N_\text{QP} \Gamma_\text{L}$ ($N_\text{QP} \Gamma_\text{R}$). The total current reads then 
\begin{equation}
I/e = \Gamma_{L} \sum_{N_\text{QP}\text{even}} P_{N_\text{QP}} N_\text{QP} + \Gamma_{R} \sum_{N_\text{QP}\text{odd}} P_{N_\text{QP}} N_\text{QP}.
\end{equation}
We now sum Eq. \eqref{eq:selfcalibrating-rate-for-nqp} over the even values of $N_\text{QP}$:
\begin{align}
0 &= -\Gamma_{\text{pb}} \sum_{N_\text{QP}\text{even}} P_{N_\text{QP}} + \Gamma_{\text{pb}} \sum_{N_\text{QP}\text{even}} P_{N_\text{QP}-2} - (\Gamma_\text{L} + \Gamma_\text{R}) \sum_{N_\text{QP}\text{even}} N_\text{QP} P_{N_\text{QP}} + (\Gamma_\text{L} + \Gamma_\text{R}) \sum_{N_\text{QP}\text{odd}} N_\text{QP} P_{N_\text{QP}}, %\\
%0 &= -\Gamma_{\text{pb}} \sum_{N_\text{QP}\text{even}} P_{N_\text{QP}} + \Gamma_{\text{pb}} \sum_{N_\text{QP}\text{even}} P_{N_\text{QP}-2} - (\Gamma_\text{L} + \Gamma_\text{R}) \sum_{N_\text{QP}\text{even}} N_\text{QP} P_{N_\text{QP}} + (\Gamma_\text{L} + \Gamma_\text{R}) \sum_{N_\text{QP}\text{even}} (N_\text{QP}+1) P_{N_\text{QP}+1}, %\\
%0 &= - (\Gamma_\text{L} + \Gamma_\text{R}) \sum_{N_\text{QP}\text{even}} N_\text{QP} P_{N_\text{QP}} + (\Gamma_\text{L} + \Gamma_\text{R}) \sum_{N_\text{QP}\text{odd}} N_\text{QP} P_{N_\text{QP}} \\
\end{align}
which leads to 
$\sum_{N_\text{QP}\text{even}} N_\text{QP} P_{N_\text{QP}} = \sum_{N_\text{QP}\text{odd}} N_\text{QP} P_{N_\text{QP}}$.
Now, as $\sum_{N_\text{QP}\text{even}} N_\text{QP} P_{N_\text{QP}} + \sum_{N_\text{QP}\text{odd}} N_\text{QP} P_{N_\text{QP}} = \sum_{N_\text{QP}} N_\text{QP} P_{N_\text{QP}} = \langle N_\text{QP} \rangle$ by definition, we obtain
\begin{equation}
I/e = \Gamma_\text{L} \langle N_\text{QP} \rangle / 2 + \Gamma_\text{R} \langle N_\text{QP} \rangle / 2 % = (\Gamma_\text{L} + \Gamma_\text{R}) \langle N_\text{QP} \rangle / 2
= \Gamma_{\text{pb}}.
\end{equation}

\subsection{Limitation of response due to recombination}

In the general case with non-negligible recombination rates, the quasiparticle population on the island satisfies
\begin{equation}
\frac{d}{dt} \langle N_\text{QP} \rangle = -2 \Gamma_\text{R} \langle N_\text{QP} \rangle ^2 - 2 \Gamma_\text{tunn} \langle N_\text{QP} \rangle +2 \Gamma_\text{pb}.
\end{equation}
The prefactors are $\Gamma_\text{R} = \Sigma \Delta^2 / [12 \zeta(5) D(E_\text{F})^2 k_B^5 \mathcal{V}]$ \cite{maisi2013excitation} and $\Gamma_\text{tunn} = (e^2 R_T \mathcal{V} D(E_\text{F}))^{-1}$ \cite{saira2012vanishing}. Here, $\zeta$ is the Riemann zeta function, the density of states $D(E_\text{F}) = 2.15 \times 10^{47}$~J$^{-1}$~m$^{-3}$ \cite{lerch2005quantum}, and $R_T$ is the resistance of the two junctions of the device in series.
The first and third factors of two are due to quasiparticles being created and recombining in pairs, while the second is due to the quasiparticles relaxing through the two junctions: as in the main text, $\Gamma_\text{tunn}$ is the relaxation rate through a single junction.
We can neglect the recombination term if $\langle N_\text{QP} \rangle \Gamma_\text{R} \ll \Gamma_{\text{tunn}}$, which is satisfied if 
\begin{equation}
\Gamma_\text{pb} \ll \Gamma_\text{tunn}^2/ \Gamma_\text{R}
% \ll \frac{ (e^2 R_T \mathcal{V} D(E_\text{F}))^{-2} }{ \Sigma \Delta^2  [12 \zeta(5) D(E_\text{F})^2 k_B^5 \mathcal{V}]^{-1} } \\
%& = \frac{1 }{ [e^2 R_T \mathcal{V} D(E_\text{F})]^2 } \frac{12 \zeta(5) D(E_\text{F})^2 k_B^5 \mathcal{V}}{\Sigma \Delta^2} \\
%&= \frac{12 \zeta(5) D(E_\text{F})^2 k_B^5 \mathcal{V}}{\Sigma \Delta^2 e^4 R_T^2 \mathcal{V}^2 D(E_\text{F})^2 } \\
= \frac{12 \zeta(5)  k_B^5}{\Sigma \Delta^2 e^4 R_T^2 \mathcal{V}} .
\end{equation}

\subsection{Thresholds for self-calibrating equation}

\begin{figure}
    \centering
    \includegraphics{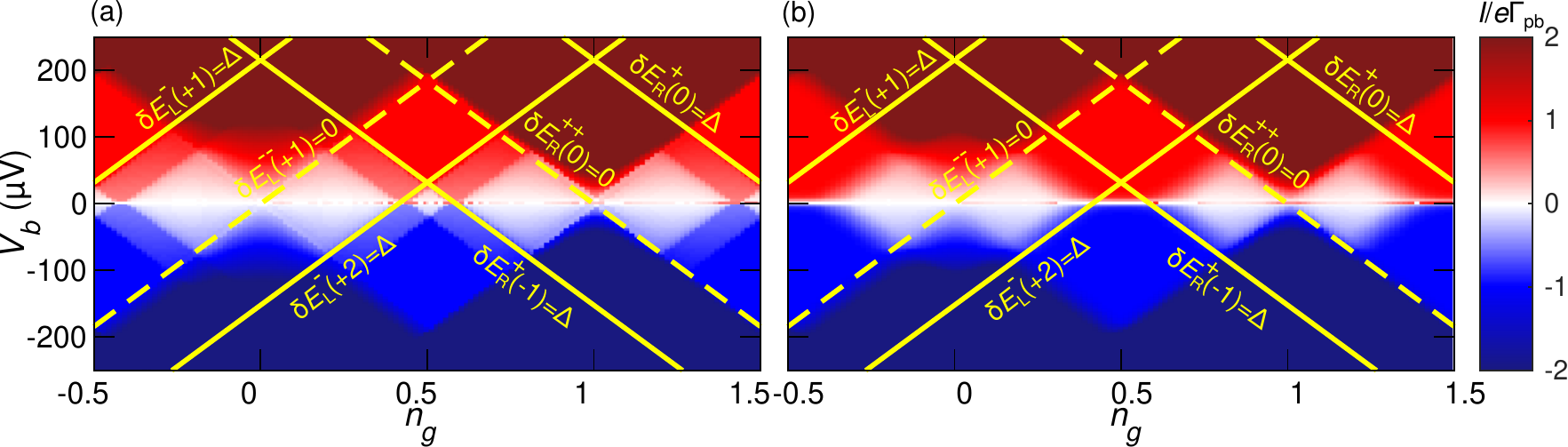}
    \caption{Simulated current through the detector with $E_C = 0.46\Delta$ and thresholds for self-calibrating operation arising from single-electron (solid lines) or Andreev (dashed lines) tunneling, calculated in (a) at $T_N = 5$ mK (essentially zero temperature) and (b) at $T_N = 42$ mK.}
    \label{fig:thresholds}
\end{figure}

The thresholds for the selfcalibrating operation can be derived from considering the energy gains for adding (+) or removing (-) an electron from the island with $N$ electrons initially through the left (L) or right (R) junctions, which include the contribution from the charging energy $E_C(N-n_g)^2$ and the voltage source $V_b$: 
\begin{align} \label{eq:de-L}
& \delta E^\pm_L(N)  = 2 E_C (\mp(N-n_g)-\frac{1}{2}) \mp eV_b/2 \\ 
\text{and } & \delta E^\pm_R(N)  = 2 E_C (\mp(N-n_g)-\frac{1}{2}) \pm eV_b/2. \label{eq:de-R}
\end{align}
Here we assume a bias voltage $V_b/2 >0$ connected to the left lead and $-V_b/2$ to the right.
Now, at zero temperature of the normal metal leads, if the energy gain satisfies $-\Delta < \delta E < \Delta$, transitions removing a quasiparticle can happen at the rate $\Gamma_{\text{tunn}} N_{\text{QP}}$ where $N_{\text{QP}}$ is the number of quasiparticles. If $\delta E > \Delta$, transitions adding a quasiparticle into the island are energetically allowed, and occur roughly at an "ohmic" rate $(\delta E - \Delta) / (e R_T)$, which is several orders of magnitude higher for typical device parameters than the rate for removing a quasiparticle. 

The self-calibrating operation occurs when the following conditions are satisfied:
\begin{itemize} 
\item The transitions $N = 0 \rightarrow +1$ and $N= +1 \rightarrow 0$ can remove a quasiparticle through either junction (recall that quasiparticles may tunnel out as both electrons and holes), but cannot add a quasiparticle: $-\Delta < \delta E_L^+(0), \delta E_R^+(0), \delta E_R^-(1), \delta E_R^-(+1)  < \Delta$. This condition ensures that the quasiparticle population satisfies Eq. (2) of the main text. 
\item The transition $N=+1 \rightarrow +2$ ($N=0 \rightarrow -1$) can remove but not add a quasiparticle through the right (left) junction: $-\Delta < \delta E_R^+(1), \delta E_L^-(0) < \Delta$. This condition guarantees that the first step of the current-carrying cycles occur with a rate $\langle N_\text{QP} \rangle \Gamma_\text{tunn}$; 
\item The transition $N=2 \rightarrow 1$ ($-1 \rightarrow 0$) can add a quasiparticle to the island through the left (right) tunnel junction, but not through the right (left) junction: $\delta E_L^-(2), \delta E_R^+(-1) > \Delta$, but  $\delta E_L^-(2), \delta E_R^+(-1) < \Delta$. This guarantees that the second step of the current-carrying cycles is fast compared to the first step and actually transports an electron through the entire device.
\end{itemize}

These conditions lead to a set of constraints for $n_g$ and $V_b$ for given values of $E_C$ and $\Delta$. If $E_C>\Delta$, the constraints are not satisfied simultaneously. For $0.5\Delta < E_C < \Delta$, the tightest constraints are by the conditions that the quasiparticles can tunnel out also against the bias between the states $N=0,+1$: $\delta E_L^+(0), \delta E_R^-(1) > -\Delta$, and that quasiparticles can tunnel out to the excited states in the direction of the bias: $\delta E_L^-(0), \delta E_R^+(1) > -\Delta$. For $E_C < 0.5\Delta$, the tightest constraints are instead the ones which guarantee the fast return from the excited charge states $\delta E_L^-(2), \delta E_R^+(-1) > \Delta$, and the constraint that tunneling events in the direction of the bias between $N=0,+1$ cannot add a quasiparticle: $\delta E_L^-(+1), \delta E_R^+(0) < \Delta$. 

At small $R_T$, additional constraints are due to the fact that Andreev tunneling should not be energetically allowed. The energy gains in Andreev tunneling $\delta E_{L/R}^{\pm \pm}$ are double the single electron gains in Eqs. \eqref{eq:de-L}, \eqref{eq:de-R} because two electrons are transported. This leads to the additional conditions that two-electron tunneling in the direction of the bias should not be energetically allowed, $\delta E_L^{--}(+1) < 0$ and $\delta E_L^{++}(1) < 0$.

These thresholds for $E_C = 0.46\Delta$ are shown superimposed to the simulated characteristics in Fig. \ref{fig:thresholds}, calculated in panel (a) at $T_N = 5$ mK and (b) at the temperature $T_N = 42$ mK corresponding to the experiment. 
At a finite temperature of the normal metal leads, the plateaus of the self-calibrating operating mode are shifted, essentially because adding a quasiparticle becomes possible already at energy gains $\delta E \gtrsim \Delta - k_B T$. Nevertheless, there is a large area in the $(n_g, V_b)$ plane where the current does not depend on the parameters.

% \bibliography{../phonons}